\documentclass[twocolumn,amsmath,amssymb,graphicx]{revtex4-1}

\draft 

\usepackage{graphicx}
\usepackage{dcolumn}
\usepackage{bm}

\begin{document}


\title{An in-situ tunable radio-frequency quantum point contact}

\author{T. M\"uller}
\email{thommuel@phys.ethz.ch}
\author{B. K\"ung}
\author{P. Studerus}%
\author{S. Hellm\"uller}
\author{K. Ensslin}%
\author{T. Ihn}
\affiliation{Solid State Physics Laboratory, ETH Z\"urich, 8093
Z\"urich, Switzerland}
\author{M. Reinwald}
\affiliation{Institut f\"ur Experimentelle und Angewandte Physik,
Universit\"at Regensburg, Germany}
\author{W. Wegscheider}
\affiliation{Solid State Physics Laboratory, ETH Z\"urich, 8093
Z\"urich, Switzerland}
\date{\today}

\begin{abstract}
Incorporating a variable capacitance diode into a radio-frequency
matching circuit allows us to in-situ tune the resonance frequency
of an RF quantum point contact, increasing the versatility of the
latter as a fast charge sensor of a proximal quantum circuit. The
performance of this method is compared in detail to conventional
low-frequency charge detection. The approach is also applicable to
other RF-detection schemes, such as RF-SET circuits.
\end{abstract}

\maketitle

Embedding single electron transistors
(SETs)~\cite{schoelkopf:98,*lu:03}, quantum point contacts
(QPCs)~\cite{qin:06,mueller:07,reilly:07,*barthel:09,cassidy:07},
and quantum dots (QDs)~\cite{barthel:10} into radio-frequency (RF)
matching circuits has become a successful technique for fast and
sensitive charge read-out of quantum dot circuits. The large
measurement bandwidths of these methods potentially allow for
probing processes at timescales beyond the scope of conventional
charge sensing, and, more straightforwardly, drastically reduce the
time required for ``standard'' measurements.

While lumped element matching networks are being used as a reliable
method for achieving accurate matching, their components are up to
now static and do not allow for fine-tuning once the system is cold
for measurement. As the circuit parameters are different at low
compared to room temperatures, good matching at room temperature
will generally not lead to the same resonance quality after cooling
down. Hence it can be a complex and time-consuming endeavor to
obtain high-quality matching at low temperatures for new samples.
Using a variable capacitance diode as a capacitor in series to a
parallel $LCR$-circuit~\cite{Note:CLCR} enables us to in-situ tune
the resonant frequency and hereby also the QPC conductance for which
optimal matching is obtained by applying a voltage across the
diode~\cite{Note:Diode}. Notably, if the measurement setup suffers
from standing waves, we can set the resonance frequency to an
anti-node, avoiding signal loss due to destructive interference.

\begin{figure}
\includegraphics{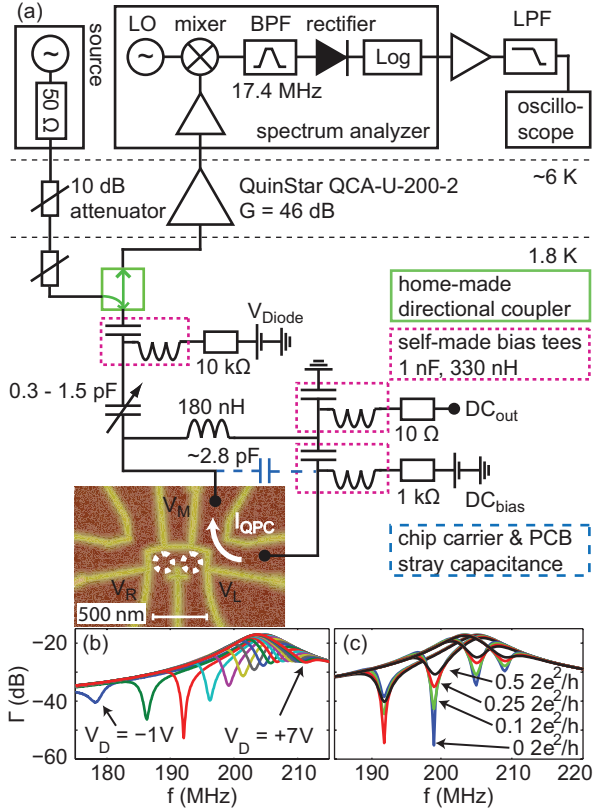}
\caption{\label{fig:setup}(Color online) (a) Experimental setup used
for time-resolved measurements. The variable capacitance diode in
series to the $RLC$ circuit allows for in-situ tunability. The
quantum point contact, defined by atomic force microscopy
lithography, is sensitive to charges in the nearby double quantum
dot. (b) Tuning of the resonance frequency by applying voltages in
steps of 0.5 V to the variable capacitance diode at fixed QPC
conductance of $0.25\times2e^2/h$. (c) Change in reflection for
different QPC conductance values at diode voltages of 0, 1, 2.7, 5 V
(ascending resonant frequencies).}
\end{figure}

The measurement setup for our variable temperature insert at
$T=1.8~\textrm{K}$ is shown in Fig.~\ref{fig:setup}(a). A
radio-frequency signal of about 200 MHz is applied, attenuated at
low temperatures, and reflected at the lumped-element matching
network containing the QPC~\cite{Note:Socket}. The reflected voltage
is amplified by 46 dB at low temperatures using a commercial
cryogenic low noise amplifier~\cite{Note:LNA} and analyzed using a
network/spectrum analyzer~\cite{Note:RS}, offering adjustable room
temperature amplification and high-quality intermediate-frequency
filtering. Simultaneous DC measurements can be performed via
self-made bias tees on the matching chip.

A double quantum dot with integrated charge sensor (see
Fig.~\ref{fig:setup}) was fabricated on a
GaAs/Al$_{0.3}$Ga$_{0.7}$As heterostructure with a two-dimensional
electron gas 34 nm below the surface
($n\sim5\times10^{15}~\textrm{m}^{-2}$,
$\mu=40~\textrm{m}^2/\textrm{Vs}$) by atomic force microscopy
lithography~\cite{fuhrer:02}. Charging an electron into the right
dot leads to a change in QPC conductance of approximately
$0.009\times2e^2/h$. This value is unusually low for AFM lithography
designed samples, probably due to the elevated temperature and the
location where the charge detector QPC is formed.

Figure~\ref{fig:setup}(b) shows the tunability of the resonance
frequency by applying a voltage to the variable capacitance diode.
With voltages in the range of -1 to +7 V \cite{Note6} we shift the
resonance frequency by more than 30 MHz. From one curve to the next,
the voltage is increased by 0.5 V, with the QPC conductance fixed at
$0.25\times2e^2/h$. In Fig.~\ref{fig:setup}(c), the QPC's
conductance is taking the values 0 (blue), 0.1 (green), 0.25 (red),
and $0.5\times2e^2/h$ (black) for different resonant frequencies
(given by $V_D=0,~1,~2.7,~5$ V). For $V_D\geq1$ V, a closed QPC
leads to the best matching, whereas for $V_D=0$ V, minimal
reflection is obtained for $G_{QPC}=0.25\times2e^2/h$. Hence we can
not only tune the resonance frequency, but also achieve matching at
a desired QPC conductance. Figures~\ref{fig:setup}(b) and (c)
exhibit a strong background due to a standing wave between matching
circuit and cold amplifier. Destructive interference significantly
reduces the reflected power for certain frequencies.

\begin{figure}
\includegraphics{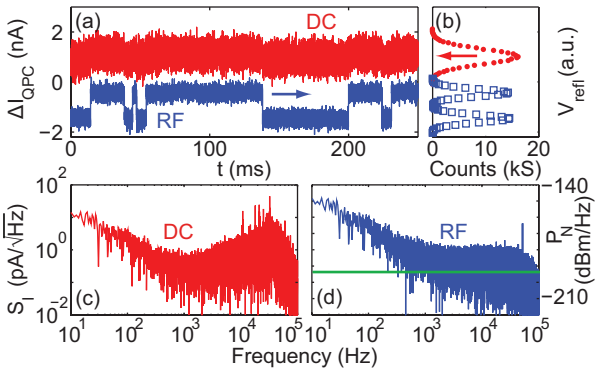}
\caption{\label{fig:Trace}(Color online) (a) Simultaneous time
traces, 8th order software filtered at 50 kHz. The feedback resistor
of the DC current-voltage converter was 1 M$\Omega$. (b) Histograms
of the time traces taken in (a). (c) and (d) Spectral densities of
the DC and RF time traces as shown in (a), low-pass filtered at 100
kHz instead of 50 kHz. The green line in (d) represents the expected
amplifier noise level at $T_N=3$ K.}
\end{figure}

A time trace of dot-lead charging is shown in
Fig.~\ref{fig:Trace}(a), with all measurement parameters optimized
as explained later-on. The DC current was measured using an
IV-converter with a feedback resistor of 1 M$\Omega$ in order to
ensure a large enough setup bandwidth. The reflected RF voltage was
rectified and logarithmically amplified (video signal) with a
spectrum analyzer. DC and RF signals have been sampled with an
oscilloscope and 8th order Bessel lowpass filtered at 50 kHz by
software. While the RF signal clearly exhibits steps whenever an
electron hops into and out of the right dot, the DC signal is
completely drowned in noise, the positions of the charging events
can only be guessed.

The histograms of these traces lead to the distributions given in
Fig.~\ref{fig:Trace}(b). While the histogram of the RF signal
reveals two distinct peaks separated by $\Delta V$, the width of the
DC current distribution is much larger than the difference of the
peak positions $\Delta I$.

Figures~\ref{fig:Trace}(c,d) show the spectral densities of the DC
and RF time traces in (a), lowpass filtered at 100 kHz. The decrease
up to frequencies of 1 kHz in both spectra originates from the
random telegraph signal itself. Beyond that, the RF spectrum is
essentially flat, while the DC spectrum exhibits a significant
increase for frequencies above a few kHz. This so-called capacitive
noise gain stems from the unavoidable capacitance of the sample
cables ($\lesssim3$ nF) and can only be reduced by putting the
IV-converter or an FET-based amplifier closer to the
sample~\cite{vink:07,hayashi:09}. For frequencies above 50 kHz, the
DC spectrum decreases due to a lowpass filter set by the feedback
resistor and its shunt capacitor, introduced to cut off the
capacitive noise gain, in combination with the finite gain-bandwidth
product of the operation amplifier. Increasing the feedback resistor
to 10 M$\Omega$ decreases this value to approximately 10 kHz, but
slightly improves the noise level. At a bandwidth of 50 kHz, the RF
setup clearly performs better, and the higher the potential
bandwidth (i.e. the larger the QPC signal) the more favorable the RF
technique will turn out to be.

\begin{figure}
\includegraphics{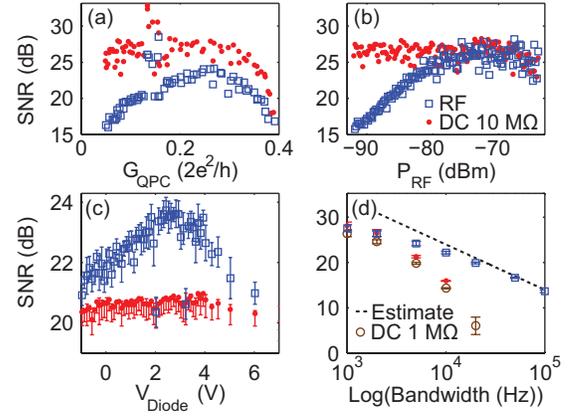}
\caption{\label{fig:SNR} (Color online) Comparison between
performance of RF reflection (blue squares) and conventional DC (red
dots) techniques. Signal to noise ratios (SNR) are given in
logarithmic power scale. With the exception of the bandwidth
dependent data in (d), all raw data has been 8th order lowpass
filtered at 1 kHz by software prior to analysis. (a) RF and DC SNR
as a function of QPC conductance at fixed diode voltage of 2.7 V.
The RF input power was set to -80 dBm, the DC bias to 250
$\mu\textrm{V}$. (b) Influence of incident RF power on DC and RF SNR
at constant DC bias of 250 $\mu\textrm{V}$ and optimal QPC
conductance of $0.27\times2e^2/h$. (c) SNR for different diode
voltages at optimal QPC values ($G=0.27\times2e^2/h$,
$P_{RF}=-75~\textrm{dBm}$, $V_{bias}=300~\mu\textrm{V}$). (d) Using
all previously determined optimal values, the frequency of the
software low-pass filter is changed to estimate the limits of both
techniques. The dashed line shows the expected linear decay of the
SNR with bandwidth.}
\end{figure}

In order to find the optimal read-out parameters for both,
conventional and high-frequency techniques, the signal to noise
ratio (SNR) of a trace of charging events (see
Fig.~\ref{fig:Trace}(a) and (b)) is determined for different
conditions. The SNR is given in dB, as the logarithm of the signal
power divided by the squared standard deviation of the histogram
peaks $\textrm{SNR}~\big(\textrm{dB}\big)=20\log\big(\Delta
I_{DC}/\sigma_I\big)$ and $20\log\big(\Delta V_{RF}/\sigma_V\big),$
respectively. Reliable counting of charging
events~\cite{gustavsson:06} using a simple threshold requires a SNR
of $\sim17$ dB ($\Delta I/\sigma_I\sim7$). A more involved,
iterative method, can lower this number
drastically~\cite{yuzhelevski:00}. Unless mentioned, the time traces
were lowpass filtered at 1 kHz to obtain a large enough SNR.

Figure~\ref{fig:SNR}(a) shows the SNR measured simultaneously with
DC (red dots) and RF (blue squares) detection for changing QPC
conductance. A DC bias voltage of 250 $\mu\textrm{V}$ and an RF
power of -80 dBm were applied, both small enough not to
significantly influence the other detection method. The diode
voltage was set to 2.7 V, fixing the resonant frequency at roughly
205 MHz. For both signals, there is a maximum slightly below
$0.3\times2e^2/h$, with the RF values decreasing faster at low
conductance. At $G_{\textrm{QPC}}\sim0.15\times2e^2/h$ there is an
additional occupancy level (probably defect charging) leading to an
overestimation of the signal height and hence too large SNR.

Choosing the optimal QPC conductance and varying the applied RF
power yields Fig.~\ref{fig:SNR}(b). The SNR increases for larger
input power, saturating above $P_{RF}=-75$ dBm, where heating sets
in, flattening out the QPC curve. From a similar analysis of DC bias
voltage (not shown), we found the threshold of the SNR saturation to
be 300 $\mu\textrm{V}$, indicating that -75 dBm corresponds to
roughly 300 $\mu$V.

Keeping the optimal QPC conductance and setting the RF and DC bias
to the values mentioned above, the influence of the diode voltage
(resonance frequency) was investigated (Fig~\ref{fig:SNR}(c)). Each
point is the average over 10 independent time traces. While the DC
SNR is unaffected, the RF SNR exhibits a maximum for voltages
between 2 and 3 V, with resonant frequencies slightly above 200 MHz.
This maximum does not occur at the best matching setting, but at the
anti-node of the standing wave (see Fig.~\ref{fig:setup}). For other
frequencies, a significant part of the signal cancels out. If the
input of the cryogenic amplifier would be perfectly matched, or if
one could mount a matched circulator~\cite{Note7} between resonant
circuit and amplifier, a significantly reduced standing wave is to
be expected and the highest SNR would be obtained for best matching.
The input power is fixed for all diode capacitances, leading to
different voltage drops over the QPC. By monitoring the QPC current,
we can ensure that the QPC's power dissipation does not
significantly change in our range of diode voltages. We observe a
slight increase ($\sim1~\%$) of current at the anti-node, though. As
the input power limit was determined for a resonance frequency at
the anti-node~\cite{Note8}, we are led to conclude that removing the
standing wave will increase the SNR ratio at optimal matching by 10
dB, the height of the standing wave.

With all parameters optimized, we varied the software low pass
filter frequency in order to find the maximum measurement bandwidth
achievable with both methods. The result is shown
in~Fig.~\ref{fig:SNR}(d). For low bandwidth (few kHz), DC - with 1
M$\Omega$ (brown circles) as well as 10 M$\Omega$ feedback
resistance - and RF perform comparably. Capacitive noise gain
deteriorates the DC SNR for larger bandwidth. An I/V-converter
feedback resistor of 10 M$\Omega$ inherently limits the DC bandwidth
to approximately 10 kHz. Lowering this value to 1 M$\Omega$
increases this cutoff to 50 kHz at the cost of a slight loss in SNR
even at low bandwidth. As the RF noise spectrum is flat above a few
kHz, the expected linear decrease of SNR with increasing bandwidth
(dashed line) can be observed. For a detection bandwidth of 50 kHz,
the RF SNR is already higher by more than a factor 10. From the
decrease of the SNR at high frequencies we can estimate the maximal
bandwidth for a SNR of unity (0 dB) to be 2.5 MHz, yielding a charge
sensitivity of $\sim6\times10^{-4}~e/\sqrt{\textrm{Hz}}$, comparable
to values reported by other groups using fast QPC charge
sensors~\cite{reilly:07,*barthel:09,cassidy:07,vink:07}. The charge
sensitivity with the DC technique is
$\sim6\times10^{-3}~e/\sqrt{\textrm{Hz}}$ - an order of magnitude
lower. As observed by others, lowering the temperature considerably
increases the sensitivity at low source-drain
bias~\cite{cassidy:07}. As charge sensitivity scales linearly with
the coupling of the dot to the QPC, using a self-aligned charge
read-out QPC on an InAs nanowire quantum dot~\cite{shorubalko:08}
should boost the sensitivity by a factor of $\sim7$ and therefore
the single shot measurement bandwidth by almost a factor 50.

In conclusion we have demonstrated in-situ tunability of a
lumped-element matched RF QPC. Its performance was compared to
conventional charge sensing after both methods were optimized with
respect to the QPC's conductance, RF power and DC voltage, as well
as resonance frequency (varactor diode capacitance) of the matching
circuit. A good charge sensitivity of
$\sim6\times10^{-4}~e/\sqrt{\textrm{Hz}}$ was achieved in spite of
the elevated temperature of 2 K. Reduction of standing waves in the
experimental setup as well as increasing the dot-to-QPC coupling can
further improve this value.

We thank P. Leek and A. Wallraff for technical advice and are
grateful to S. Gustavsson for the measurement software. Our research
was funded by the Swiss National Science Foundation (SNF).


\end{document}